\documentclass{elsart}
\usepackage{amssymb}

\usepackage{graphicx}
\usepackage{setspace}
\usepackage{bm}

\newcommand{\be}{\begin{equation}}
\newcommand{\ee}{\end{equation}}
\newcommand{\bd}{\begin{displaymath}}
\newcommand{\ed}{\end{displaymath}}
\newcommand{\ba}{\begin{array}}
\newcommand{\ea}{\end{array}}
\newcommand{\bq}{\begin{eqnarray}}
\newcommand{\eq}{\end{eqnarray}}

\def\thesection{\Roman{section}}
\usepackage{doublespace}
\begin{document}
\begin{frontmatter}

\title{FERM3D: A finite element R-matrix electron molecule scattering code}
\author{Stefano Tonzani}
\address{JILA and Department of Chemistry,  University of Colorado,
Boulder, Colorado 80309-0440}
\ead{tonzani@colorado.edu}

\date{\today}

\begin{abstract}
FERM3D is a three-dimensional finite element program,
for the elastic scattering of a low energy electron from a general polyatomic molecule,
which is converted to a potential scattering problem.  The code is based on tricubic
polynomials in spherical coordinates. 
The electron-molecule interaction is treated as a sum of three terms:
electrostatic, exchange. and
polarisation. 
The electrostatic term can be  extracted directly from ab initio codes
({\sc{GAUSSIAN 98}} in the work described here),
while the exchange term is approximated using a local density
functional. A local polarisation potential based on density functional theory
[C. Lee, W. Yang and R. G. Parr, { Phys. Rev. B} {37},  
(1988) 785] 
describes the
long range attraction to the molecular target induced by the scattering
electron. Photoionisation calculations are also possible and illustrated in the
present work. The generality and simplicity of the approach is important in
extending electron-scattering calculations to more complex targets than it is
possible with other methods.

\end{abstract}

\begin{keyword}
Electron scattering \sep Polyatomic molecules \sep Finite Elements \sep
R-matrix \sep DNA radiation damage \sep Photoionisation \sep Density functional
theory \sep Polyatomic molecules.
\PACS 31.15.Ew \sep 34.80.-i \sep 34.80.Bm
\end{keyword}

\end{frontmatter}

\section{Program summary}
{\em Title of program}: FERM3D\\
{\em Catalogue identifier}:\\ 
 {\em Program Summary URL}: 
http://fermion.colorado.edu/\textasciitilde tonzani/Software/rmatrix/Software.html \\
 {\em Program obtainable from}: CPC Program Library, Queen's University
of Belfast, N. Ireland\\
 {\em Computer for which the program is designed and others on which it
has been tested}: Intel Xeon, AMD Opteron 64 bit, Compaq Alpha\\
 {\em Operating systems or monitors under which the program has been
tested}: HP Tru64 Unix v5.1, Red Hat Linux Enterprise 3\\
 {\em Programming language used}: Fortran 90\\
 {\em Memory required to execute with typical data}:  900MB (neutral
CO$_2$), 2.3GB (ionic CO$_2$), 1.4GB (benzene) \\
 {\em No. of bits in a word}: 32 \\
 {\em No. of processors used}: 1 \\
 {\em Has the code been vectorized?}: No\\
 {\em No. of bytes in distributed program, including test data, etc.}:  \\
 {\em Distribution format}: Tar gzip file \\
 {\em CPC Program library subprograms used}: ADDA, ACDP \\
 {\em Nature of physical problem}: Scattering of an electron from a complex polyatomic
molecular target. \\
 {\em Method of solution}: Solution of a partial differential equation using a finite
element basis, and direct sparse linear solvers. \\
 {\em Restrictions on the complexity of the problem}: Memory constraints. \\
 {\em Typical running time}: 2 hours. \\
 {\em Unusual features of the program}:
\begin{itemize}
\item Very extensive use of memory
\item  requires installation of Lapack, Blas, a direct sparse solver library (SuperLU, freely
available, or Pardiso, which requires a license, but is free of charge for
academic use), and optionally the Cernlib and Arpack libraries, freely available.
\item  requires input from quantum chemistry programs (Gaussian, Molpro or PC Gamess).\\
\end{itemize}

\section{Introduction}
Electron-initiated chemical processes are recognized as increasingly pivotal
in many areas such as cell radiation damage \cite{Sanche:DNA}, the treatment of
wastes, 
interstellar/atmospheric processes \cite{McCall:H3}. 

The advances both in
theory \cite{Resc:99} and experiment \cite{Allan:JPB03,Burrow:PRL04}
in recent years have been significant. However, there is a lack of theoretical
work on large molecular targets, because of the difficulties inherent with
such calculations, which are memory and time consuming. 
Calculations have been mostly limited thus far either to small molecules 
or to large but highly symmetric molecules like C$_{60}$ or SF$_6$ in their equilibrium 
configurations\cite{Gianturco:c60,Gianturco:SF6}.  Recently, however,  calculations for
the 13 atom nonsymmetric tetrahydrofuran (THF) molecule have been performed
using both the R-matrix UK code \cite{Gorfinkiel:JPB06} and the Kohn complex
code \cite{Trevisan_Orel_THF:06}, while calculations for uracil have been
performed with the Schwinger variational code \cite{Sanchez_McKoy:05}. Gianturco and co-workers have
also performed calculations on uracil \cite{Gianturco:uracil_JCP04}, using a similar approximation to what we
describe here.
We present here a detailed description of a code we have
developed to deal with more complex nonsymmetric molecules; examples of  the
capabilities of this approach have already been shown in previous works
\cite{Tonzani:JCP05,Tonzani:JCP06}, and here we want to stress the practical
aspects of  the code and describe  its structure in detail. The main
advantage of our approach, which is based on a grid basis set and the R-matrix
method, is its simplicity and adaptability with changing molecular targets.
Because the code does not use Configuration Interaction (CI) expansions or
require fitting of the continuum basis set, it is  
is very simple to use. Nevertheless it has good predictive capability, in the limit of the
approximations used, and it does not rely on adjustable parameters.
Symmetry properties of the targets are not used, thus the code can be applied
to large molecular targets even in unsymmetric
configurations, such as an SF$_{6}$ molecule distorted from its
equilibrium octahedral geometry. This feature allows us to take into account vibrational
processes and vibronic interactions.

\section{Theory}
\label{sec:Theory}
Since the theory behind our approach has been described in Refs.
{\cite{Tonzani:JCP05,Tonzani:JCP06}}, we will limit
ourselves here to a few generalities. The code solves a one-particle Schr\"{o}dinger
equation for an incoming electron in the field of a model potential that
describes the interaction of the electron with the target molecule as:
\be
\label{1-body:eqn}
(-\frac{1}{2}\nabla^{2}+V_{s}+V_{ex}+V_{pol})\phi_{0}(\vec{r})= E\phi_{0}(\vec{r}),
\ee
where $V_{s}$ is the local electrostatic potential for the interaction of the
incoming electron with the target electron density and the nuclei, and $V_{ex}$
is the nonlocal exchange potential arising from the antisymmetrisation of
the wavefunction. $V_{pol}$ is a polarisation potential (also in
principle nonlocal) that describes the relaxation of the target under the
influence of the incoming electron, and $\phi_{0}(\vec{r})$ is the continuum
electron wavefunction.

The main approximations we make
are: (1) static exchange (which means that we only consider the
ground state surface of the target in the wavefunction expansion
\cite{Morr_Coll:PRA78}), (2) representation of the nonlocal
exchange potential by a local form using a local density approximation (LDA)
and (3) adoption of a local polarisation potential based on Density
Functional Theory (DFT) instead of the
full nonlocal many body polarisation. 
These approximations essentially
reduce the initial many-body process to a potential scattering problem and they
also enable
the use of this code for relatively large targets.

The local density approximation leads to an exchange potential of the
form:
\be
\label{Pot:exch}
V_{ex}(\vec{r}) = -\frac{2}{\pi} k_{F} F(k_{F},E,I),
\ee
where $k_{F}$ is the local Fermi momentum,
\be
k_{F}(\vec{r})=[3 \pi ^{2} \rho(\vec{r})]^{1/3},
\ee
and $F$ is a functional of the energy and the local
density $\rho(\vec{r})$ (through the local Fermi momentum). The  functional
form used for $F$, called the Hara
exchange \cite{Hara:69},  has been extensively employed in continuum-states
calculations, it is energy-dependent, and depends on the ionisation
potential ($I$ in Eq. \ref{Pot:exch}) of the molecule.
The basis set used is composed of a cross product of cubic finite elements in
all three dimensions.
An introduction to these basis sets and their practical implementation can be seen in Refs.
\cite{FEMBEM:notes,Bathe:book}. Our basis set guarantees functional and
derivative continuity at any point, and also guarantees
flexibility in the choice of the grid points which is fundamental for highly
anisotropic objects, such as polyatomic molecules. It also generates highly
sparse matrices,
which is important for large molecules.

The
scattering observables are obtained from
Eq. \ref{1-body:eqn} using the R-matrix method \cite{Greene:rev96}. 
In essence, we solve the Schr\"{o}dinger equation inside a a spherical box of radius
$R_{0}$,  using a variational principle for the logarithmic
derivative of the wavefunction:
\be
\label{var_princ:eqn}
{b\equiv - \frac{\partial{\ln {(r\Psi)}}}{\partial r} = 2 \frac
{\int_V {\Psi^{*}(E-\hat{H} -\hat{L}) \Psi dV}}{\int_V{\Psi^{*}
\delta(r-r_0)\Psi dV}}},
\ee
where $\hat{L}$ is the Bloch operator \cite{Greene:rev96}.
By expanding in some basis set inside the box, Eq. \ref{var_princ:eqn} is
reduced to the
solution of a large eigenvalue problem and then,  through basis set
partitioning \cite{Tonzani:JCP05,Greene:rev96}, to a large linear system plus a
small eigenvalue problem.
At this point, it is worthwhile to comment on the difference between our method
of solution and the usual R-matrix calculations based on diagonalisation of the
Hamiltonian. One of the strengths of the R-matrix method is the possibility of
solving for all energies with just one diagonalisation of the Hamiltonian, but
this is a weakness when treating large systems because all the eigenvalues of
the matrix are required, therefore a huge amount of space is needed to run the
calculation. In our method we have to solve a linear system at each energy, but
in turn we gain from the sparsity of the matrices, since the linear system
never requires more than, say, 5\% of the memory required to store the whole
eigenvector matrix, which would be necessary in the usual R-matrix formulation.
Competing solutions to address this problem in R-matrix theory are being
developed through partitioning schemes \cite{Tennyson_partit:JPB04}.
Outside the sphere, the long-range potentials are accounted for by matching
them to appropriate long-range wavefunctions such as Bessel functions for neutral
molecules or Coulomb functions for ionic molecules. We do not account for
multipole terms higher than the dipole outside the box, although this would be
a straightforward modification.

\subsection{DFT polarisation potential}
\label{sec:DFT}
 As shown in Ref. \cite{Lane:rev80}, the
polarisation-correlation contribution  (basically the distortion-relaxation
effect on the molecule generated by the incoming electron) is extremely important for an accurate
description of the scattering process. 
We have recently added the capability of using a parameter-free
correlation-polarisation potential\cite{Gianturco:PRA93,Colle_Salvetti:JCP83},
which is based on 
DFT ideas.
The long-range part of this potential
can be thought of as 
a simple multipole expansion, of which we retain only the dipole terms: 
\be
\label{polar:potential}
V_{pol} = -\frac{1}{2 r^{4}}(\alpha_{0}+\alpha_{2} P_{2}(\cos{\theta})),
\ee
where $P_{2}(\cos{\theta})$ is a Legendre polynomial, and  $\alpha_{0}$ and $\alpha_{2}$ are the totally symmetric and nontotally
symmetric components of the polarisability tensor, calculated from ab
initio electronic structure codes.
In the area where the electronic density of the target is not negligible,
this potential is nonlocal; however the interaction can be approximated as a
local potential. Different forms have been suggested in the literature; the one
we use is based on DFT (specifically on the Lee-Yang-Parr potential of
Ref. \cite{Lee_Yang_Parr:PRB88}) and has been proven to give
reliable results in the work of Gianturco and coworkers \cite{Gianturco:c60}.
This form makes use of the electron density, its gradient, and Laplacian, all
of which have to be calculated for each target. The short- and long-range potentials are
matched unambiguously at their first crossing point, which is angle dependent.
The final potential is continuous but not smooth.

\section{Target molecular structure}
\label{sec:Structure}
To extract the molecular electron density, gradient, Laplacian, and the
electrostatic potential needed to construct the scattering potentials, we use
{\sc{GAUSSIAN 98}} (version g98-A9) \cite{GAUSSIAN}, but {\sc{MOLPRO}} \cite{MOLPRO} has
the same capabilities and exploits the same file formats, as does
{\sc{PC-GAMESS}} \cite{PC-GAMESS}. The latter is also freely available.  We usually perform
structure calculations, as described in Ref. {\cite{Tonzani:JCP05}}, at the Hartree-Fock (HF) or configuration interaction
with single and double excitations level for the target molecule. We use a cubic grid with
120 points or more per dimension. 
The R-matrix sphere has to be completely contained in the grid cube, so the
R-matrix radius has to be smaller than the grid outermost extent, in each
dimension. The potential
calculation is not very expensive. On an Intel Xeon 2.4Ghz machine it typically
requires forty minutes for a
large molecule like the DNA base guanine, 
at the HF level with a 6-31G* basis set. The files generated from
these calculations are rather large (at least 20 MB each), so it is impossible
to include them in the program distribution. However, we include Gaussian scripts 
that allow the user to generate these files. 
The large files so generated and necessary to run the examples given in this article can be found at
the FERM3D website \cite{FERM3D:website}.

In the calculation of the potentials,
particular care has to be taken that no grid point falls exactly on an atom
position, since interpolation errors could ensue. This particular problem only
occurs with
atoms at the origin of the Cartesian axes, and can be avoided by a slight 
shift in the origin of the ab initio grid. 
If FERM3D detects this condition, it will stop. An error will
also occur if the outermost extent of the grid is smaller than the R-matrix radius.
It is important to notice that the Coulomb potential is not interpolated as
such, but first the singularities at the atom positions are subtracted and then
the interpolation is performed. In the calculation of the matrix elements
the singularities are added again; since we use Gauss-Legendre integration, no
Coulomb singularity is ever sampled.

\section{Photoionisation}
\label{sec:Photoionisation}
The relationship between electron scattering and photoionisation is a 
close one \cite{Dill_Dehmer:JCP74}. The scattering wavefunction relative to a photoionisation process is essentially
the time-reverse of the one for the scattering of an electron with a positive ion.
Thus it is useful to examine the resonant structures of the
electron with the ionic state, since photoionisation allows to distinguish
between different initial wavefunction symmetries.
The photoionisation cross sections can be
obtained starting from the dipole matrix elements,
\be
\label{eq:dipole_length}
d^{(-)}_{0, l'm'}=< \psi^{(-)}_{l'm'} \mid \hat{\epsilon} \cdot  \vec{r} \mid \psi_{0} >,
\ee
using the length form of the dipole,
where $\hat{\epsilon}$ is the polarisation of the incident photon, $\psi_{0}$
the initial wavefunction (the orbital from which the electron is ionized), and 
$\psi^{(-)}_{l'm'}$ the final state of the continuum electron. This is formed
from R-matrix eigenstates $\psi_{l'm'}$ using
\be
\psi^{(-)}_{l'm'}=\psi_{l'm'}(I-iJ)^{-1},
\ee
where $I$ and $J$ are matrices generated from the Wronskians of the solutions
inside and outside the R-matrix box,
as described in Eqs. 2-31 and 2.41 of Ref. \cite{Greene:rev96}, generating a wavefunction with
incoming wave boundary conditions:
\be
\psi^{(-)}_{l'm'}=\sum_{lm} Y_{lm}(\theta,\phi) (\frac{1}{i \sqrt{2}}f^{+}_{l}
\delta_{ll'} \delta_{mm'}
- \frac{1}{i \sqrt{2}}f^{-}_{l} {S}_{lm, l'm'}^{\dagger} ),
\ee
where $f^{\pm}$ are incoming and outgoing wave solutions, respectively, and
$S^{\dagger}$ is the hermitian conjugate of the scattering matrix.
The total cross section can be easily calculated from the $d^{(-)}_{0 l'm'}$
matrix elements by performing a rotational average expressing the dipole
operator as
\be
\hat{\epsilon}  \cdot \vec{r}= \sum_{m_{\gamma}} {\frac{4 \pi}{3}}r
Y_{1m_{\gamma}}(\theta, \phi) Y^{*}_{1m_{\gamma}}(\theta', \phi').
\ee
where the second spherical harmonic refers to the
molecular orientations, and $m_{\gamma}$ gives the projection of the polarisation
vector on the molecular $z$ axis. 
After angular integration, the final cross section is:
\be
\label{eq:photoion_sigma}
\sigma=N \frac{4 \pi \omega}{3} \sum_{l'm'} d^{(-)}_{0, l'm'} {d^{(-)*
}_{0, l'm'}}
\ee
where $N$ is the number of degenerate electrons that can be knocked out by the
photon. For example, $N=4$ for a fully occupied $\pi$ orbital in CO$_{2}$. The
cross section ouptut from the code has to be multiplied by the orbital
occupation number.
As an alternative to Eq. \ref{eq:dipole_length}, the velocity form of the dipole
matrix element,
\be
d^{(-)}_{0, l'm'}=\frac{1}{\omega}< \psi^{(-)}_{l'm'} \mid \hat{\epsilon} \cdot \vec{\nabla} \mid \psi_{0} >,
\ee
can be used in the code, although this will be slower because of the need to
differentiate many continuum wavefunctions. In fact, in the case of many initial
wavefunctions, it is too expensive to store all of their
gradients, so the final wavefunctions have to be differentiated at each energy point.

\section{Post processing}
\label{sec:post_proc}
The post-processing stage allows us to calculate dipole effects
on the total elastic scattering cross section \cite{Tonzani:JCP06}, even though
the effects of higher partial waves are not included through a Born closure-type
formula \cite{Tonzani:JCP06,Gianturco_Stoecklin:JPhysB96}. To use the post-processing
code, it is necessary to have the Cernlib library installed, which is  freely available from CERN
\cite{cernlib}. This version of the code is not set up to handle Coulomb +
dipole external fields or negative energy calculations of quantum defects.
This code also allows us to calculate
differential cross sections, for molecules with zero permanent dipole
moment. To do this we applied formulas 46-47 in Ref.
\cite{Dill_Dehmer:JCP74}.  An example of this calculation appears in Fig.
\ref{fig:DCS_CO2} for CO$_{2}$. The results can be compared
with those of Refs. \cite{Resc:99,Gianturco_Stoecklin:JPhysB96,Morrison:PRA77,Brunger:JPB99}.  

An additional feature is the
possibility of plotting wavefunction maps corresponding to the eigenvectors
associated with the largest eigenvalue of the time-delay matrix,
\be
\label{eq:timedel}
Q=iS\frac{dS^{\dagger}}{dE},
\ee
which corresponds, on resonance,  to the channel that experiences the longest
time delay. This term constitutes the main contribution to the resonance and is 
useful in identifying the symmetry and nodal structures of the resonant
wavefunction, as shown for DNA bases in Ref. \cite{Tonzani:JCP06}.

\section{Calculation details}
\label{sec:Details}
For the examples included with the code distribution (CO$_{2}$ and benzene as
neutral targets and CO$_{2}^{+}$ as ionic
target), the calculations are not too expensive, the maximum dimension of the
matrices involved being about 50000-70000. When stored in a sparse format, these
matrices can be
stored in a 32 bit machine without any parallelisation. Note that the CO$_{2}$
photoionization calculation requires 2.3GB of memory, because of the
transformations required to generate the dipole matrix elements. For more
complicated targets, such as DNA bases, it is necessary either to use 64
bit machines or else to parallelize the linear solver. For the latter case, we have
experimented with different parallel solvers, SuperLU and WSMP (Watson sparse
matrix package). The first is freely distributed, but has problems working with
Intel Fortran compilers, the second has license fees. We have
retained the option of using SuperLU \cite{SuperLU} only in sequential mode in the present
version of the code, since it is freely available also for commercial
applications. 
Iterative solvers were considered, but because the
linear system is often ill-conditioned, requiring many iterations to
reach convergence (if it converges at all) this option has also been eliminated. The current version of
the code uses mainly the Pardiso solver \cite{pardiso},  which is freely
distributed for academic purposes by the University of
Basel, but it requires a license for commercial applications. This library is
included in the Intel Fortran mathematical libraries. In general the user will
have to refer to the Pardiso website \cite{pardiso} for instructions on how to get a license and use this
library.
An important difference between the SuperLU and Pardiso solvers is that the
former (at least the version we have tested) does not support a symmetric mode,
therefore it requires twice as much memory. We have also tested our code more
extensively with Pardiso.
The libraries required by the code are shown in Tab. \ref{tab:libraries}

As mentioned earlier, the memory requirements for 
calculations on large molecules are quite large, around 6 gigabytes for calculations on purines and
pyrimidines. The maximum matrix dimensions can be on the order of 250000, and the
time needed for setting them up is around 90 minutes. Finally, each energy
point requires about 30 minutes to solve the linear system and calculate cross
sections and wavefunctions. 

It is necessary to stress the importance of choosing an adequate grid,
especially around the nuclei where the electron density varies very fast, to
get converged results. The
reader is urged to read our previous published works on this code, in
particular Sec. IIE of Ref. \cite{Tonzani:JCP05} and Sec. IID of Ref.
\cite{Tonzani:JCP06} for details on the grids and the integration. In general,
the grid spacing close to the nuclei should be comparable to the K-shell of the
atom, in order to describe the wavefunction correctly, as in Fig. 2 of Ref.
\cite{Tonzani:JCP06}.

\section{Code structure}
\label{sec:Code_Structure}
\begin{figure}[p]
\newpage
\centerline{\includegraphics[width=10.0cm]{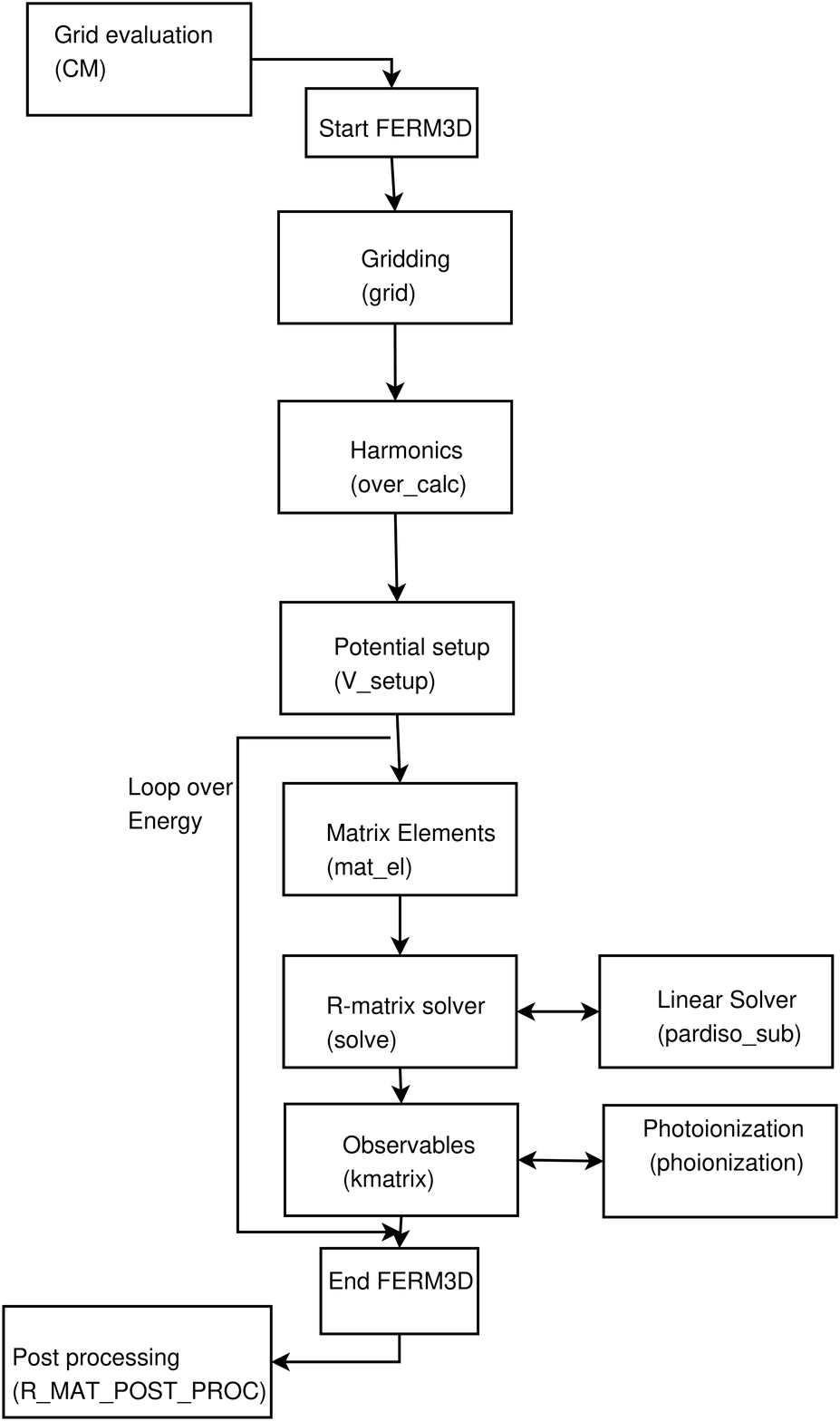}}

\begin{spacing}{2}
\newpage
\caption{Code diagram. Each box shows the main function of a
program/module, with  the program names in capital letters, and in
lowercase the module/subroutine names enclosed in parentheses.
}\label{fig:Diagram}
\end{spacing}
\end{figure}

The FERM3D package is made up of three codes: (1) the general scattering code
(FERM3D.x), (2) an automatic grid utility (CM.x), and (3) a post-processing
module (R\_mat\_post\_proc.x) to extract differential cross sections and wavefunction maps
\cite{Tonzani:JCP06}. The instructions for the compilation are inside the
package. For each run, a set of four files from ab initio codes are required:
electrostatic potential, electron density, norm and Laplacian. A script for
Gaussian to calculate these files is also included in the distribution. 
From the quantum chemistry codes are also required: the
spherical and nonspherical polarisabilities, the ionisation potential of the
target and, for post-processing, the dipole moment magnitude and its components along the
Cartesian axes. In the case of an ionic target, a photoionisation calculation
can be performed, requiring the initial orbitals on a grid in the same format as for
the potentials, and the ionisation potentials of these orbitals.
Then the
file input\_control.dat should be set up (see Section \ref{sec:Input}).
Afterwards, CM.x should be run to get the
grid file. At this point the scattering calculation can be performed. When the
calculation is completed,
the post-processing code can be run to get supplementary information on the
system as specified in Section \ref{sec:post_proc}. 
As shown in Fig. \ref{fig:Diagram}, the main code has the following structure :
grid calculation (performed in subroutine {\bf {grid}}), numerical
harmonics calculations ({\bf{over\_calc}}), potential setup
({\bf{V\_setup}}),  matrix elements setup ({\bf{mat\_el}}). Next comes the linear
system solution ({\bf{lin\_rmatsparse}}) and the calculation of cross sections
({\bf{kmatrix}}). Finally, an optional  
photoionisation calculation ({\bf{photoionisation}}) can be run.
Some subroutines in the code
have been taken from Numerical Recipes \cite{Num_Rec:book}, others from the
Computer Physics Communications library \cite{CPC:link}, and others from other
sources. All sources are cited in detail in the code.
 
As already stated, a number of third party libraries are used in this
code, some necessary, some optional. Blas/Lapack are needed throughout the
code, and a direct sparse solver (SuperLU in routine "dSLUsolve" or Pardiso in
"pardiso\_sub") is also needed, since the
large matrices generated by this code cannot be handled using a dense solver,
even for a small target molecule.
Optional libraries are Arpack\cite{Arpack:website}, used to calculate surface harmonics in routine
"over\_calc" together with the sparse solvers (otherwise a dense solver is
used, for small/medium targets there is no difference in performance) and
Cernlib, used in the post-processing utility. This is shown in Tab.
\ref{tab:libraries}.

\begin{table}
\newpage
\begin{tabular}{|l|c|c|c|} \hline
Libraries & Intel & Alpha & SuperLU \\ \hline
\emph{Fortran Compiler}   & ifc (Intel, v9.0.027) & f90 (HP, v.V5.4-1283)& f90 \\ \hline
\emph{C Compiler}   &  & & gcc(v.3.0.1) \\ \hline
\emph{General}   & Lapack(v.3.0) & Lapack &  Lapack \\ \hline
\emph{Eigenvalue}   & Arpack(v.96) &  &  \\ \hline
\emph{Linear system}   & Pardiso(v.3.0) & Pardiso & SuperLU(v.2.0) \\ \hline
\emph{Post processing}   & Cernlib(v.2004) & &  \\ \hline

\end {tabular}
\begin{spacing}{2}
\newpage
\caption{Libraries needed to link the three different sample versions of the code
given with the package, for alpha and Intel platforms. The Intel version shown
here also uses the optional Arpack and Cernlib libraries, which can be avoided.
in parentheses is indicated the version of the libraries used here. Essentially
all possible library combinations can be utilized, i.e. using SuperLU on Intel
platforms and so on. To
interface the code with SuperLU 3.0 the bridge routine dSLUsolve.c will have to be modified.}
\label{tab:libraries}
\end{spacing}
\end{table}

\section{Input description}
\label{sec:Input}
Here we summarize the input file structure for the codes. In the input
files reproduced below, the first entry (or two in the case of two variables
per line) for each line corresponds to a sample value for the variable, e.g.
for the CO$_2$ examples we show here. On the right side, there are comments
including the name of the variable in the code and some brief
descriptions. 
The input to CM.x is the file
CM.dat, which specifies the maximum spacing per dimension. The output file is
fort.16 which should be copied into input\_grid.dat.  The structure of CM.dat
is:
\begin{verbatim}
12.d0         ! R0 = R-matrix box size
0.125         ! Delta_theta_max = Delta theta max
0.125         ! Delta_phi_max = Delta phi max
0.125         ! Delta_r_max = Delta R max (internal)
.5d0          ! Delta_external_radius = R max (external)
\end{verbatim}
where the first entry defines the R-matrix box size R0. The box should contain the
molecule and be completely contained in the cubical grid files output from ab initio programs
for the potentials. A value of R0 from 10 to 14 Bohrs is usually appropriate for most of the
molecules we have worked with. The other inputs define a maximum size of the finite
element sectors in $\theta$, $\phi$ (both in units of $\pi$), and $r$. In the case
of the radial dimension, two areas are defined, internal ($r<$r\_cutoff which
is defined below) and external. All the variables in this file are real and double precision.
Also this code takes as input some quantities
from input\_control.dat, which is described below.

The grid spacing is fundamental for the convergence of the calculation. In
simple molecules like CO$_2$, our gridding utility can easily do a good job of
ensuring convergence while guaranteeing a low memory usage. For larger
molecules some manual adjusting of the grid, especially around the nuclei, is
often required.
It is necessary for the convergence of the integrals to have a grid point at
the position of each nucleus. For a correct representation of the
wavefunction, it is also advisable to have grid points near the K-shell radius
of the nuclei.
CM.x constructs grids taking these factors into account, but verification of
the convergence of the results (for example increasing the number of points in
some of the dimensions by 50\% or so) is critical for validation of the
results, in other words, especially for inexperienced users that do not yet
know what grid spacings should be used in each dimension, the first run on a
molecule can give unconverged results that will not be very meaningful.
Also, we have verified time and again that the angular grid spacings are more
critical than the radial ones, since for atoms far off center of the grid,
the $\theta$ and $\phi$ angles are usually described more poorly. Usually a
spacing of $0.0625\pi$ in the angular variables outside the nuclei K-shell is fine enough for most molecules. In the
examples we give here $0.125\pi$ is enough to get satisfying convergence since
CO$_2$ is a compact target, while in benzene the atoms far from the grid center
are hydrogens.

The main code has input\_grid.dat and input\_control.dat as input files. 
Data files named pot.dat, dens.dat, grad.dat, and nabla.dat must be created
for the potential, density, norm of the gradient and Laplacian respectively,
using output from ab initio codes.
The input for the post-processing module is included in input\_control.dat. This file has the following
structure for the CO$_2$ example discussed in this paper:
\begin{verbatim}
950             ! output_file = File number for elastic cross section 
scatter   nowf  ! calculation_type, option_wf
slater    8     ! type of exchange (obsolete),max_l
ion       1.0d0 ! molecule,  charge = residual molecular charge
8.95d0          ! polar = spherical polarisability
4.595d0         ! polar2 = nonspherical polarisability
3.7d0           ! r_cutoff = cutoff radius for DFT polarisation
0.90d0          ! Energy_last_bound = Ionisation potential
0.d0            ! shift = shift on z axis
1.2d0           ! Emax = Energy max
0.0d0           ! Emin = Energy min
12              ! nEmax = number of energy points
.0d0            ! shifty = shift on y axis
.FALSE.         ! rotation = swapping of y,z axes
.TRUE.          ! option_DFT = DFT correlation
0.d0            ! shiftx = shift on x axis
no              ! restore (obsolete)
no        0.d0   0.d0  0.d0  0.d0  ! option_DCS, dipole(5)


!!!! PHOTOIONIZATION INPUT !!!!!!!!!!
3   length      ! num_initial_states = Number of initial orbitals, gauge 
pi_u_1.dat      ! orbital_name1 = Initial orbital file for photoionisation
0.71            ! Ioniz_potential = Ionisation potential for the initial state
sigma_g_4.dat
0.71
homo.dat
0.54
\end{verbatim}

Now we will describe various variables, which have not been explained
above. We indicate the name of the variable, with its type in parentheses. In
our description,
I stands for integer, R for double precision real, S for strings, all
of which are of length
10 except orbital\_name1 which can be of length 16,  Then we explain the
variable's meaning and, for
string variables, the values they can assume:\\
General part:\\
{\bf{output\_file}} = (I) redirects the output elastic cross section to file
fort.output\_file.\\ 
{\bf{calculation\_type}} = (S) defines the kind of calculation wanted, e.g. scattering
(value = scatter) or bound-state problem (value = bound) to get the bound states of
the potential.\\
{\bf{option\_wf}} = (S) sets the print switch for the scattering wavefunction,
either to print
(value = wf) or not (value = nowf) the eigenvector associated with the largest
eigenvalue of the time-delay matrix of Eq. \ref{eq:timedel}.\\
{\bf{max\_l}} = (I) maximum continuum angular momentum included (highest
partial wave included in the calculation).\\
{\bf{molecule}} = (S) defines the molecule type, which can be a neutral (value = neutral) or positive
ion (value = ion).\\
{\bf{polar, polar\_2}} = (R) represent one half of  $\alpha_{0}$ and $\alpha_{2}$ in
Eq. \ref{polar:potential} respectively.\\
{\bf{r\_cutoff}} = (R) ensures the uniqueness of the crossing point between the DFT and
asymptotic polarisation potentials.  This variable should be set to a value roughly two atomic units
larger than the radius of the atom farthest from the grid center.\\
{\bf{Energy\_last\_bound}} = (R) defines the ionisation potential of the
target, and enters the calculation
from Eq. \ref{Pot:exch}.\\
{\bf{shift, shifty, shiftx}} = (R) coordinates translation for the whole molecule on the
three Cartesian axes (largely unnecessary, but can be used to try to minimise the
memory requirements).\\
{\bf{option\_DFT}} = (S) offers a choice between DFT correlation (value = DFT, strongly
recommended) and
semiempirical (value different from DFT), as in Ref. \cite{Tonzani:JCP05}.\\
{\bf{option\_DCS}} = (S) creates a switch in post-processing for the differential cross section, to print
it (value = yes) or not (value = no).\\
{\bf{dipole}} = (R) constituted by an array of four numbers, the first of which is the absolute value of the dipole,
the other three its x,y,z components.
After these variables there are three blank lines for eventual future development.\\
Photoionisation part (only required for ions):\\
{\bf{num\_initial\_states}} = (I) specifies the total number of initial
orbitals to photoionise from. \\
{\bf{option\_gauge}} = (S) specifies which gauge to use
to calculate the dipole matrix element (can assume value=length, which
is the default, or value=velocity).
For each initial orbital (1 to  {\bf{num\_initial\_states}}) the filename
{\bf{orbital\_name1}}(S), which can assume any value, and the ionisation potential, {\bf{
Ioniz\_potential}}(R), of the orbital are specified. 

The code checks the correctness of the input options, and in case they are
erroneous (for example a negative polarisability), it flags an error and exits.
It is worth noticing that error checking is done on the grids as well, however
this is not completely foolproof, as some subtle errors might not be caught by
the checking done here. In these cases though, we have verified that the linear solver will exit before
completion, flagging an error.

For the examples included in the package, the input files are in the directory
"examples" for each of the given tests. The inputs and outputs are included for
each example in separate subdirectories, and the file structure is explained in
a README file.

\section{Output description}
\label{sec:Output}
\begin{figure}[p]
\newpage
\centerline{\includegraphics[width=18.0cm]{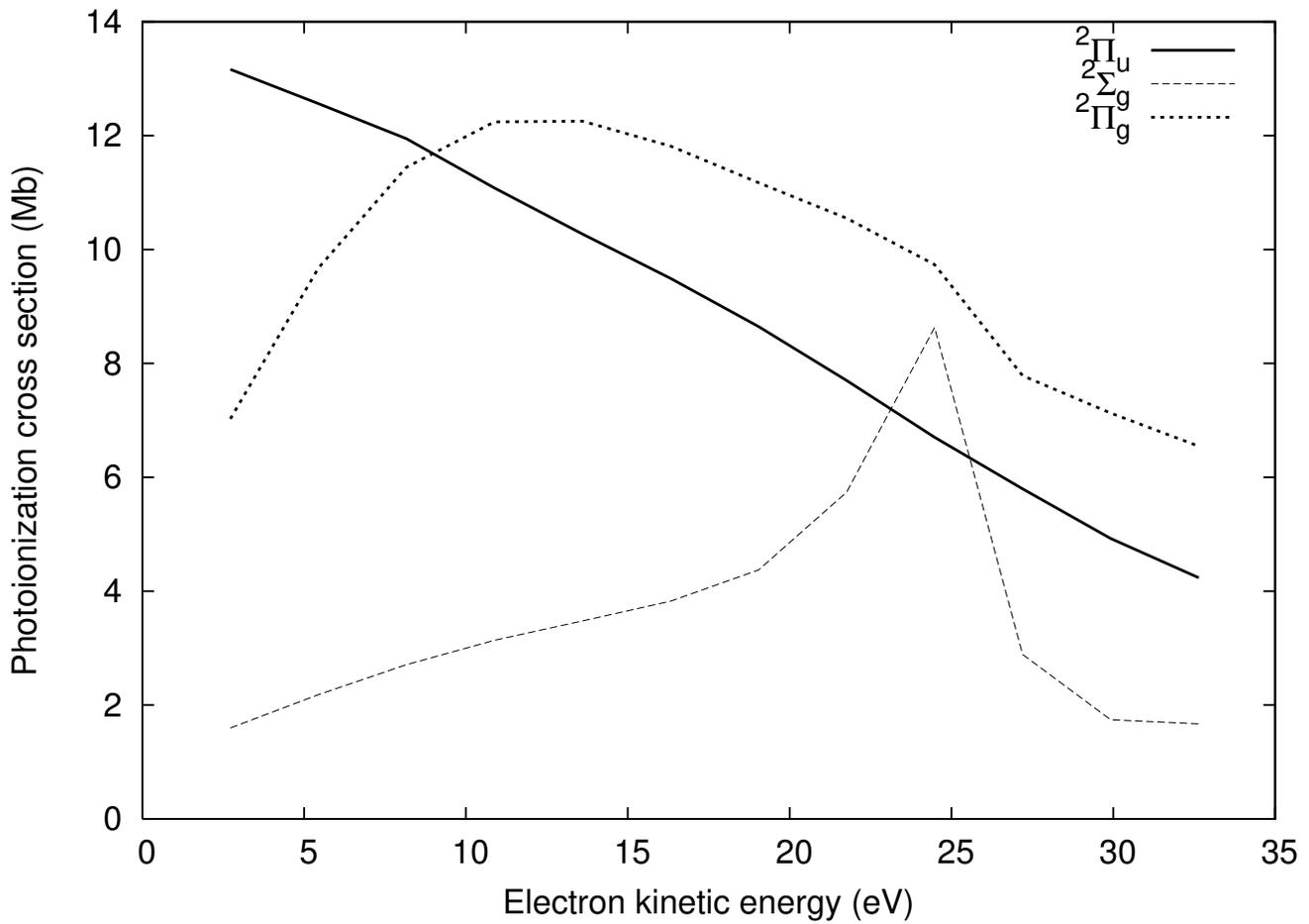}}

\begin{spacing}{2}
\newpage
\caption{CO$_{2}$ photoionisation cross section for the three highest valence
orbitals ($\pi_{u}$, $\pi_{g}$, $\sigma_{g}$), which generate the doublet final states
in the figure. Comparison with the results of Lucchese et al. \cite{Lucchese_CO2:PRA82} is
very good. In our calculation the $\sigma_{u}$ resonance is shifted roughly 1 eV
higher with respect to Ref. \cite{Lucchese_CO2:PRA82}.
}\label{fig:CO2_photoion_cross_section}
\end{spacing}
\end{figure}

\begin{figure}[p]
\newpage
\centerline{\includegraphics[width=18.0cm]{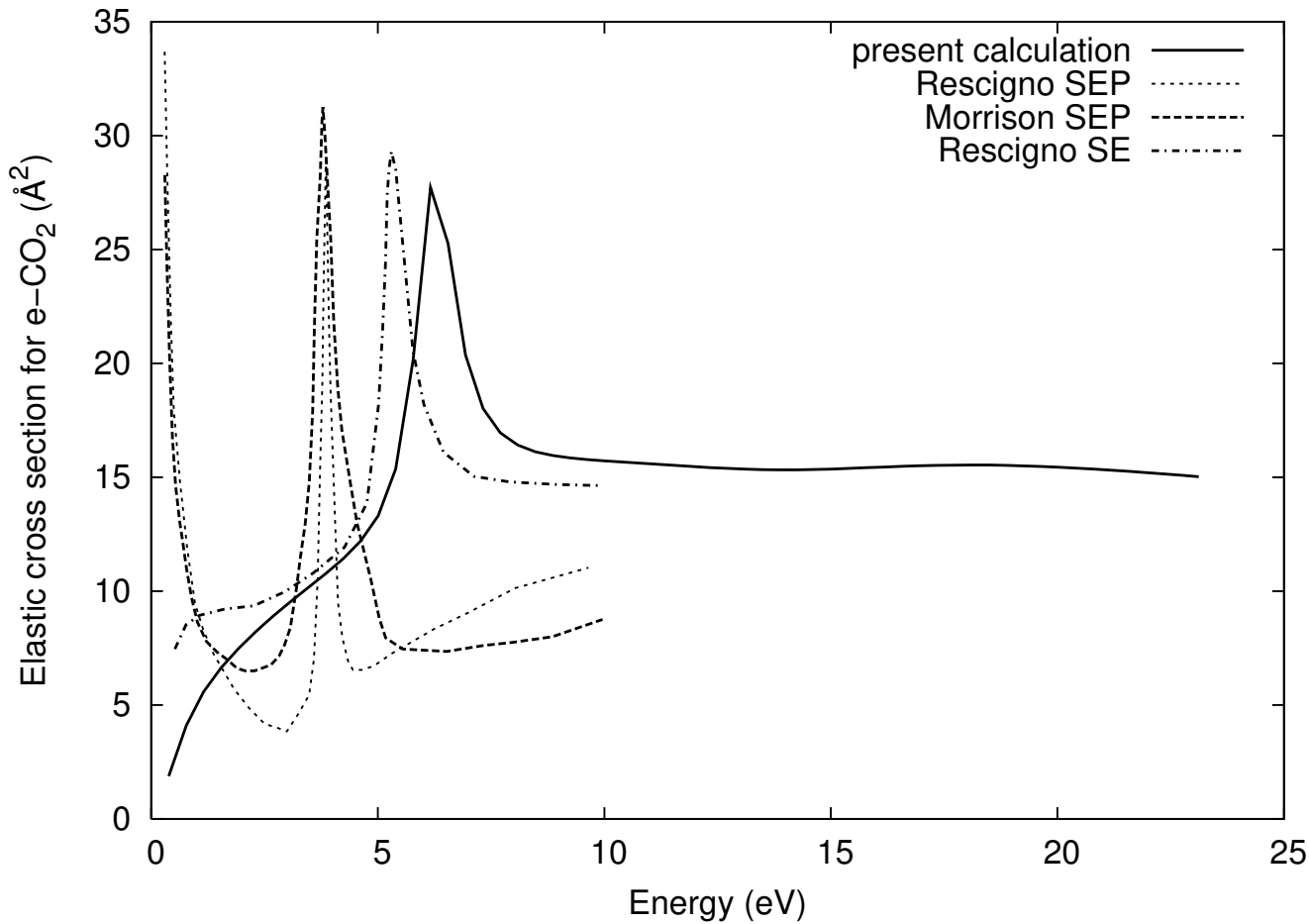}}

\begin{spacing}{2}
\newpage
\caption{CO$_{2}$ elastic cross section at low energies. The prominent
structure is the well-known $\pi_{u}$ resonance which, in experiment, has been
found at 3.8 eV. Often the resonances we calculate with
our approach are
roughly 2-2.5 eV higher if  compared to experimental values, see e.g. Fig. 4 in Ref. \cite{Tonzani:JCP05}.
Our results are compared
with those of Rescigno $et$ $al.$ \cite{Resc:99} (exact static exchange and
exchange plus polarisation) and Morrison and Collins \cite{Morr_Coll:PRA78} (model exchange plus semiempirical
polarisation). The resonance width is larger than the SEP results of Rescigno
$et$ $al.$ but comparable to the other calculations.
}\label{fig:cross_section_elastic_CO2}
\end{spacing}
\end{figure}

\begin{figure}[p]
\newpage
\centerline{\includegraphics[width=18.0cm]{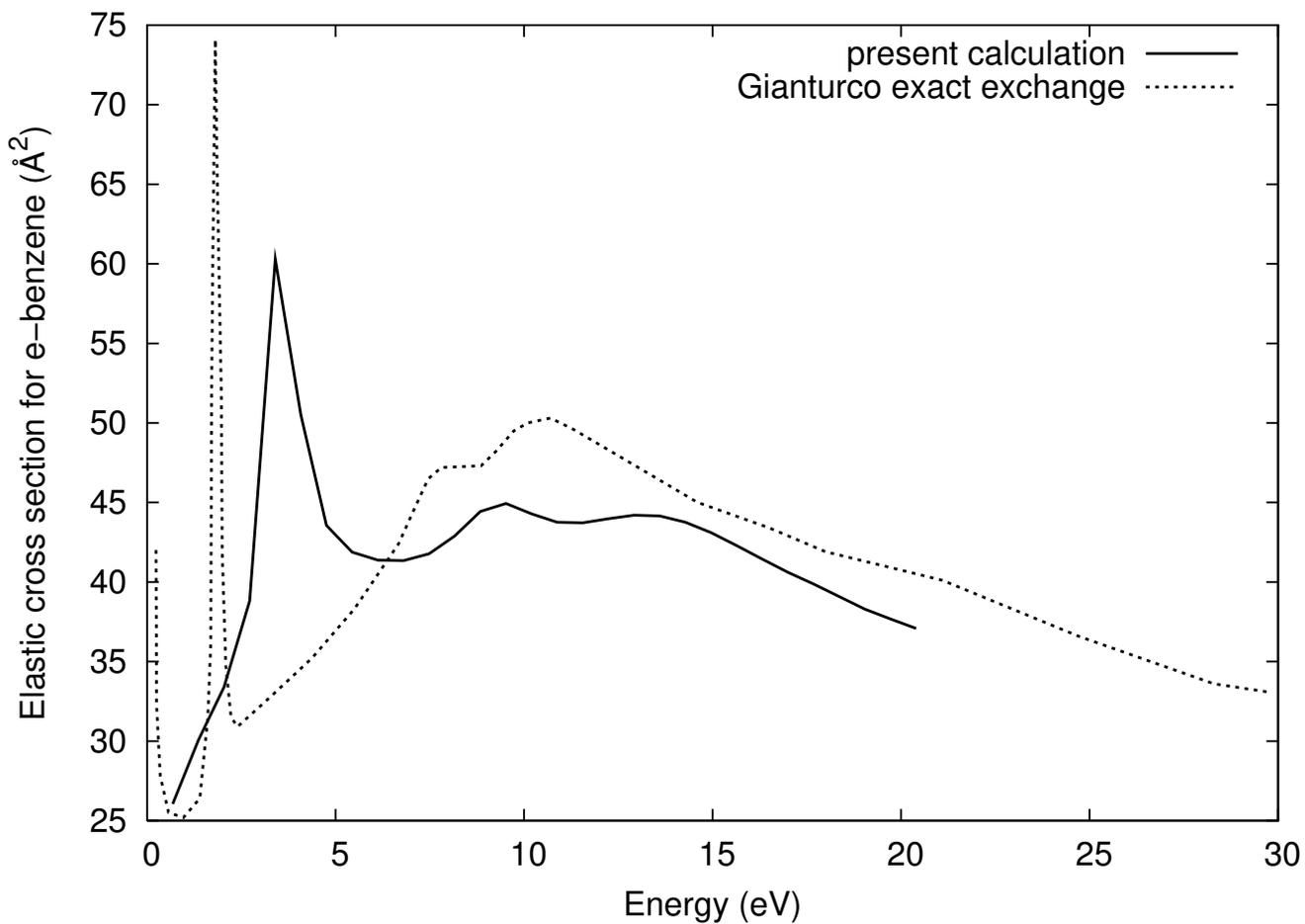}}

\begin{spacing}{2}
\newpage
\caption{Benzene elastic cross section at low energies. Comparison with the
exact exchange plus model polarization  results of Gianturco $et$ $al.$ \cite{Gianturco:benz}.  The comparison of the resonance
positions with the model exchange results of Ref. \cite{Gianturco:benz} are
given in the text. }
\label{fig:cross_section_benzene}
\end{spacing}
\end{figure}

\begin{figure}[p]
\newpage
\centerline{\includegraphics[width=18.0cm]{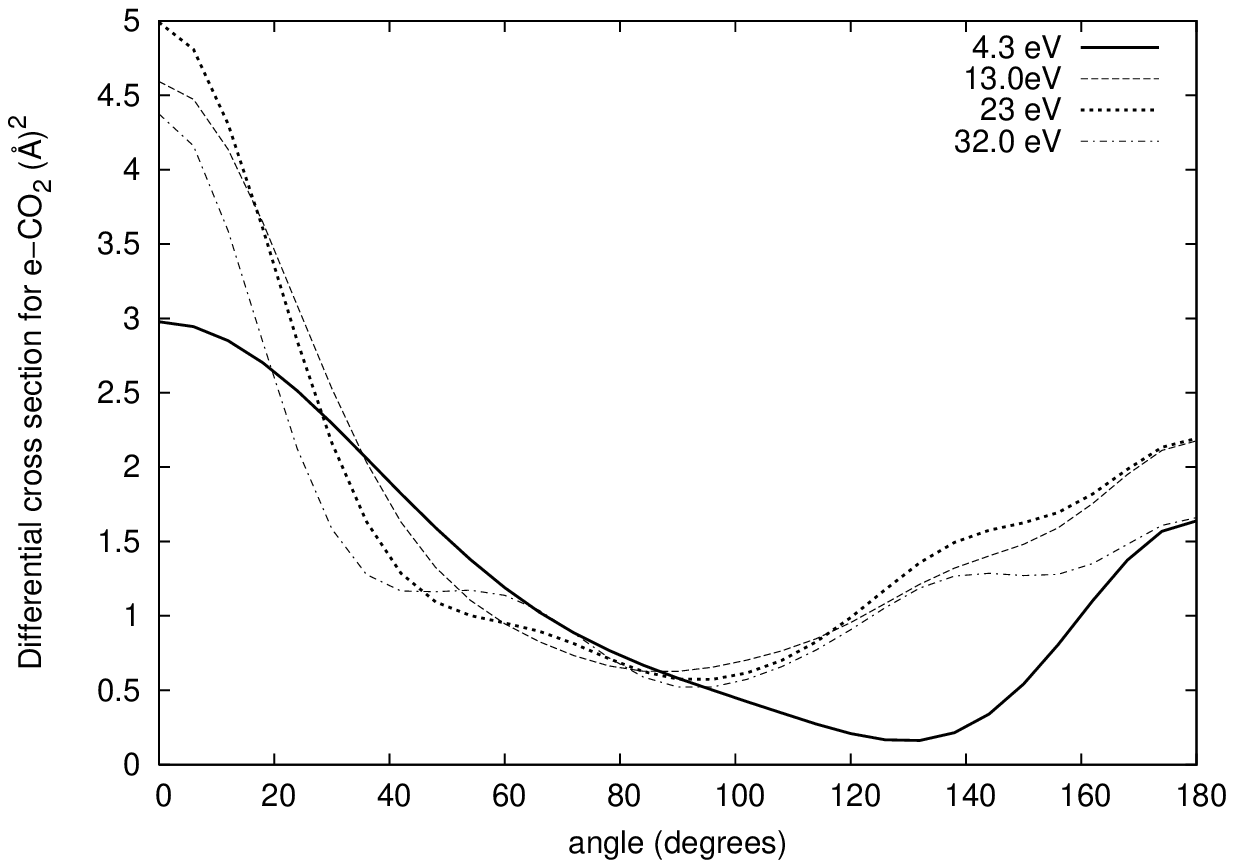}}

\begin{spacing}{2}
\newpage
\caption{CO$_{2}$ differential cross section at four selected low electron
impact energies. The results can be compared with those of Ref.
\cite{Resc:99,Gianturco_Stoecklin:JPhysB96,Morrison:PRA77,Brunger:JPB99}.
}\label{fig:DCS_CO2}
\end{spacing}
\end{figure}

Sample calculations for the photoionisation of CO$_{2}^{+}$, total elastic and
differential elastic cross sections of  
CO$_{2}$ and elastic scattering off of benzene  are illustrated in Figs.
\ref{fig:CO2_photoion_cross_section},\ref{fig:cross_section_elastic_CO2},\ref{fig:DCS_CO2},\ref{fig:cross_section_benzene}
respectively. 
The program can output total elastic cross sections (file number
set in input\_control.dat, as specified in Sec.\ref{sec:Input}), phase
shifts (phase\_shifts.out), S-matrices (fort.536), total photoionisation cross
sections from Eq. \ref{eq:photoion_sigma}
(fort.951 and following, one for each initial state orbital), the dipole matrix
elements $d^{(-)}_{\beta}$ in Eq. \ref{eq:photoion_sigma} (momenta.out). Through the post-processing modules, differential
cross sections (fort.100 and following, one for each energy point calculated),
eigenvectors of the time-delay matrix (fort.13001), elastic cross sections 
 and S-matrices with a dipole potential outside the R-matrix box
(fort.1280 and fort.538, respectively) can be output as well. Many of these
quantities are also written to standard output. 
The matrices are all written by columns, the wavefunction file contains the x,
y and z
coordinates of the point and the wavefunction value, in this order. The other
files all have the energy in the first column and the quantity of interest in the
second.

For the examples given in the package, the output files for total elastic and
photoionisation cross sections (together with the
input scripts) are listed in the directory "examples" of the tar file. The
figures included in this article were generated using the examples given.

In general, the approximations made in the model are such that the resonant
structures will appear shifted too high with respect to experimental values, in
many cases we have seen that this shift is roughly of the order of a couple of
eV, as for the DNA bases \cite{Tonzani:JCP06} and THF \cite{Tonzani:THF} and
also for smaller molecules like  carbon dioxide, benzene, as shown in Figs.
\ref{fig:cross_section_elastic_CO2} and \ref{fig:cross_section_benzene} and
SF$_6$,\cite{Tonzani:unpublished}. 
The fact that our resonances are shifted too high in energy can in turn cause
the width to be larger than the experimental one. We have verified in the past
for many systems \cite{Tonzani:JCP06,Tonzani:unpublished}
that our calculated widths are usually close to their correct values.

The model potential we use here has been used extensively in the literature, in
particular a comparison is possible with data from Gianturco and co-workers, we
will cite the values of the resonances we found for the benzene molecule, and
compare to the ones listed in Table II of Ref. \cite{Gianturco:benz}. In our
calculations the resonances are at 3.4eV, 9.2 eV, 13.2, compared to values
of 4.66, 9.02, 12.25, so the values are quite similar. Also the widths are
quite similar. It is not
possible to compare directly the cross sections because this data is not
available from Ref. \cite{Gianturco:benz}, however we will compare with their
results using correct (nonlocal) exchange, which shifts all their resonances
down by a few eV, in Fig. \ref{fig:cross_section_benzene}. It is important to notice how in this case
the higher resonances have roughly the same widths between the model potential and
exact exchange calculations, while the lowest one becomes much narrower and it
is shifted down by 3eV with respect to the others. In Ref. \cite{Tonzani:JCP05}
we have also compared to the results of Morrison and Collins
\cite{Morr_Coll:PRA78} for CO$_2$, using the same adjustable polarisation
potential used in that work. We have
shown how those results are very similar to our calculations in that case. In
Fig. \ref{fig:cross_section_elastic_CO2} we also compare our results for CO$_2$
with those of Refs. \cite{Resc:99} (exact static exchange, SE, and
exchange plus polarisation, SEP) and \cite{Morr_Coll:PRA78} (local exchange plus semiempirical
polarisation).
For larger molecules, the only comparison possible is with the results of
Gianturco $et$ $al.$ on uracil \cite{Gianturco:uracil_JCP04}, which we have given in detail in Ref. \cite{Tonzani:JCP06} and will not be repeated here.

\section{Conclusions}
We have presented the FERM3D package, a general code for performing electron
molecule scattering and photoionisation calculations. The code is very versatile and
can be used to extract, with predictive value, elastic cross sections, 
resonance structures, and photoionisation cross sections. No use is made of
symmetry, whereby the code performs
well for molecules with low symmetry, such as the DNA bases, which are difficult to
explore with more sophisticated approaches.

\section*{Acknowledgements}
We wish to thank Prof. C. H. Greene for the stimulus in undertaking this project
and the continuous help and support given along the way, and J. Phillips for
proofreading the manuscript.
This work was supported by the U.S. Department of Energy's Office of Science, by
an allocation of National Energy Research Scientific Computing Center (NERSC) resources, and by the Keck Foundation
through computational resources. 

\bibliography{paper_FEA}

\end{document}